\documentclass[aps,prd,preprintnumbers,noshowpacs,superscriptaddress,nofootinbib,notitlepage]{revtex4-1} 
\usepackage{amssymb}
\usepackage{graphicx,verbatim}
\usepackage{epstopdf}
\usepackage{slashed}

\IfFileExists{srcltx.sty}{\usepackage[active]{srcltx}}

\begin{document}

\title{Can supersymmetry breaking lead to electroweak symmetry breaking via formation of scalar bound states?}

\preprint{IPMU12-0191}

\author{John M. Cornwall}
\affiliation{Department of Physics and Astronomy, University of
California, Los Angeles, CA 90095-1547, USA}

\author{Alexander Kusenko}
\affiliation{Department of Physics and Astronomy, University of
California, Los Angeles, CA 90095-1547, USA}
\affiliation{Kavli Institute for the Physics and Mathematics of the Universe,
University of Tokyo, Kashiwa, Chiba 277-8568, Japan}

\author{Lauren Pearce}
\affiliation{Department of Physics and Astronomy, University of
California, Los Angeles, CA 90095-1547, USA}

\author{R. D.  Peccei}
\affiliation{Department of Physics and Astronomy, University of
California, Los Angeles, CA 90095-1547, USA}

\begin{abstract}

The recent discovery of the putative 125-GeV Higgs boson has motivated a number of attempts to reconcile its relatively large mass with the predictions of the minimal supersymmetric standard model (MSSM).  Some approaches invoked large trilinear supersymmetry-breaking terms $A_t$ between stops and one of the elementary  Higgs fields. We consider the possibility that electroweak symmetry breaking may be triggered by supersymmetry breaking with a large $A_t$, large enough to generate a composite field with the same quantum numbers as the Higgs boson and with a non-vanishing vacuum expectation value.  In the resulting vacuum, the usual relation between the gauge couplings and the Higgs self-coupling does not apply, and there is no reason to expect the same upper bound on the mass of the lightest Higgs boson. In a simple model where the bound state is assumed to have no mixing with the other fields, 
we calculate the critical coupling $A_t$ necessary for symmetry breaking using the lowest-order Bethe-Salpeter (BS) equation. 
Study of the BS equation is complicated by the structure of its lowest-order kernel, which is a crossed box graph, but we find 
an accurate approximation to its solution.  In a realistic model, the mixing of the bound state with the fundamental Higgs boson 
creates a symmetry-breaking seesaw.   We outline the steps toward a realistic model. 
\end{abstract}

\maketitle 

The recently discovered  125~GeV boson~\cite{arXiv:1207.7214,arXiv:1207.7235} is widely thought to be the Higgs boson.  The mass outside the range predicted by the 
simplest supersymmetric models, combined with the lack of evidence for  superpartners~\cite{:2012ew}, has encouraged a number of efforts to reconcile  
low-energy supersymmetry with a relatively heavy Higgs boson (see, e.g., Refs.~\cite{dmrs,Evans:2012hg,Yanagida:2012ef}).  The models usually assume heavy masses for the superpartners, 
as well as some novel features, for example, strong couplings in the supersymmetry breaking sector~\cite{Yanagida:2012ef}.  While many models with gauge-mediated supersymmetry breaking 
predict a small tri-linear supersymmetry breaking coupling $A$, a large value of such coupling is, in fact, helpful in raising the range of the Higgs boson masses toward 125~GeV~\cite{Okada:1990gg,dmrs}.   Large tri-linear terms can appear in gauge-mediated supersymmetry breaking models, albeit some fine-tuning of parameters may be required in a realistic model~\cite{dmrs}.

However, if the tri-linear couplings are large, the low-energy realization of supersymmetry may differ dramatically from the usual set of predictions.  It has been pointed out~\cite{kkt,gk} that the 
exchange of the (lighter) Higgs boson between  (heavier) squarks can lead to formation of bound states, resonances, and a new strongly coupled realization of the minimal supersymmetric standard model (MSSM).  Here we reconsider this possibility and will focus, in particular, on the possibility that supersymmetry breaking may trigger  electroweak symmetry breaking via formation of relativistic squark bound states  having the quantum gauge numbers of the Higgs boson.  Such new states can mix with the Higgs boson, they can acquire a vacuum expectation value (VEV), and the resulting multi-Higgs low-energy effective theory may have a very different appearance from the usual weakly coupled MSSM.  At the same time, the ultraviolet behavior of the theory is preserved, and supersymmetry provides the usual solution to the hierarchy problem.  The difference is in the low-energy effective theory, which contains different degrees of freedom: fewer squarks, and more Higgs bosons, whose VEVs produce a more complicated vacuum.   In this vacuum, the usual MSSM relations between the gauge couplings and the scalar self-coupling do not hold, and, therefore, there is no reason for the upper bound on the lightest Higgs boson to be the same as in the usual version of MSSM.

Let us consider a simplified version of MSSM, in which we will focus only on the third generation of squarks and will assume that only one tri-linear term is large: 
%
\begin{equation}
\label{lagrangian}
\mathcal{L}=A_t(\tilde{t}^{\dagger}_L\cdot\phi)\tilde{t}_R+h.c.
\label{At}
\end{equation}
where $\tilde{t}_L$ is the Y = 1/3 stop doublet under $SU(2)_L$, $\tilde{t}_R$ the Y = 4/3 stop singlet, and $\phi$ the Y = -1 Higgs doublet.  We omit writing $\phi^4$ terms.  For simplicity we assume the squarks have a common mass $M\sim$ a few TeV, considerably larger than the Higgs mass $m$.  

We have suppressed the SU(3) indices in Eq.~\ref{At}, and  will concentrate on the color-singlet bound state.  This is the only bound state that can have a mixing with the fundamental Higgs boson, and as discussed below, we expect there to be a range of parameters in which this bound state has a non-zero VEV, while all the SU(3) non-singlet bound states (which can form through the Higgs exchange as well) have a zero VEV.  This case corresponds to the standard-model-like multi-Higgs vacuum consistent with the data.

\begin{figure}[ht!]
\includegraphics[width=0.9\textwidth]{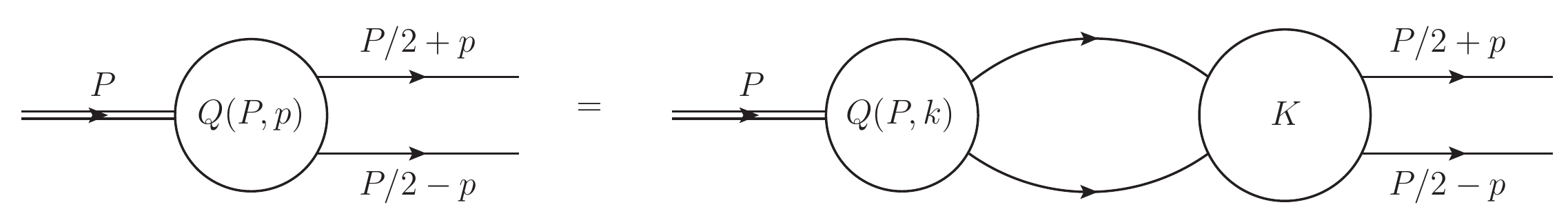}
\caption{\label{fig1-bs}  The Bethe-Salpeter equation in ``vertex" form. The double line represents the bound state $\Phi$, of momentum $P$, and the single lines represent the constituent squarks, of mass $M$.}
\end{figure}

We seek a CP+ scalar doublet with Y = 1 (the quantum number of one of the Higgs fields) arising as a $(\tilde{t}_R\tilde{t}_L^{\dagger})$ bound state described by a Euclidean BS equation as shown in Fig.~\ref{fig1-bs}.

This BS equation is in vertex form, where the internal lines represent propagators and the usual BS wave function $\Psi (P,p)$ is related to $Q(P,p)$ by
\begin{equation}
\label{wavefunct}
\left [\left (\frac{P}{2}+p \right )^2+M^2 \right ] \left [\left (\frac{P}{2}-p \right )^2+M^2 \right ]\Psi (P,p)=Q(P,p).
\end{equation}
It is useful to state the BS equation in this form because it is closely related to a gap equation whose non-trivial solution yields an estimate of how large the coupling in the kernel $K$ must be to yield a bound-state Higgs that mimics the Higgs field $\phi$.  This bound-state Higgs with symmetry breaking mixes $L$ and $R$ stops and contributes to their mass difference.

\begin{figure}[ht!]
\includegraphics[width=0.5\textwidth]{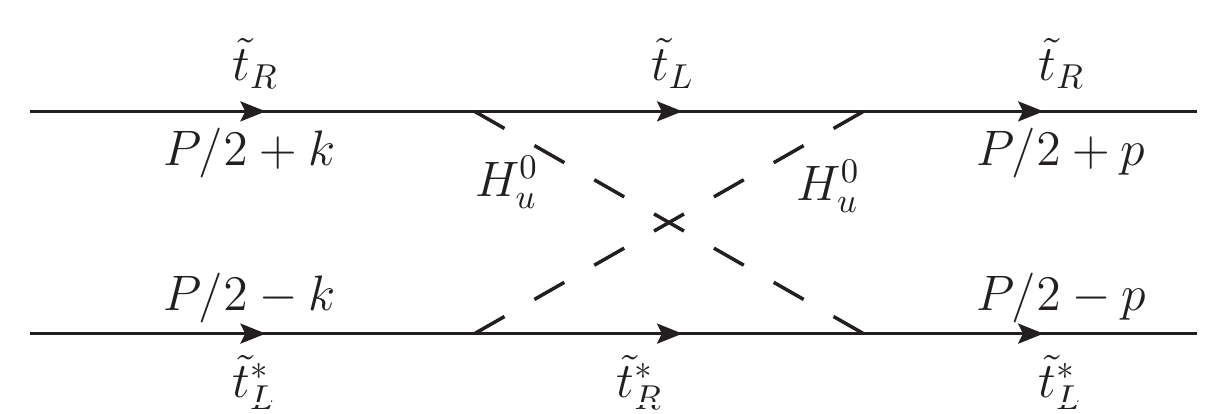}
\caption{\label{fig2-kern}  The lowest-order kernel for the Higgs-like bound state.}
\end{figure}

Let the line labeled $(P/2)+p$ represent an outgoing $\tilde{t}_R$, and the line labeled $(P/2)-p$ represent an outgoing $\tilde{t}_L^{\dagger}$.  A few minutes of drawing Feynman graphs shows that the lowest-order kernel must be a crossed box graph, as shown in Fig.~\ref{fig2-kern}.

The lines labeled $L,R$ are the stops, and the dashed lines are the Higgs fields of the MSSM.  
The Euclidean BS equation is
\begin{equation}
\label{bseq}
Q(P,p)=\frac{A_t^4}{(2\pi )^4} \int \mathrm{d}^4k\,Q(P,k)
\frac{1}{\left [ \left (\frac{1}{2}P+k \right )^2+M^2 \right ] \left [\left (\frac{1}{2}P-k \right )^2+M^2 \right ]}K(P,p,k).
\end{equation}
We have omitted writing an $SU(2)$ spinor index on $Q$ and a corresponding factor $\delta_{ij}$ on the kernel.  

The complications of this one-loop kernel have prevented us from studying this BS equation at general  momentum $P$, which would furnish a relation between the values of $P$ for which the equation is solvable and the coupling $A_t$.  Instead, we look for the value of $A_t$ at which $P=0$, corresponding to having four degenerate massless bound states. This total of four massless states corresponds to the two complex elements of the bound-state $SU(2)$ spinor. Alternatively, these states can also be considered as states of a broken $SU(2)\times U(1)$ theory: a zero-mass composite Higgs boson, plus three zero-mass Nambu-Goldstone bosons.     We are interested in Higgs bosons whose mass is comparable to $m$, the lightest Higgs boson mass in the MSSM.  By hypothesis this is close to 125 GeV, while the squark mass $M$ is much larger. Thus,  it should be a reasonable approximation to consider the bound-state Higgs as having zero mass and study the $P=0$ BS equation.  In addition, this $P=0$ BS equation can be tackled 
with decent quantitative accuracy.  From now on we use the notation $K(p,k)= K(k,p)$ for the original kernel at $P=0$,  whose bound states can only occur for specific values of $A_t$ that are eigenvalues of the BS equation.   

In the general case with $P\neq 0$ the BS wave function $\Psi (P,p)$ is the Fourier transform of the matrix element
\begin{equation}
\label{bswf}
\psi (X,x) = \langle 0|T(\tilde{t}_L^{\dagger}(x_1)\tilde{t}_R(x_2))|P \rangle
\end{equation}
with $P$ conjugate to the center-of-mass coordinate $X=(1/2)(x_1+x_2)$ and $p$ conjugate to $x_1-x_2$; the state $|P\rangle$ is the bound state.  At $P=0$ this looks like a vacuum-to-vacuum propagator $\Delta_{LR}(x_1-x_2)$, but one must be careful about what the vacuum means.  Just as in superconductivity the true vacuum is a non-perturbative construct quite different from the bare vacuum; in our case the true vacuum has matrix elements connecting $L$ and $R$ stops.  This connection comes from a symmetry-breaking order parameter that is a mass splitting  $\delta M^2(p)$   found in this $LR$ propagator, vanishing in the symmetric case, that mixes $L$ and $R$ squarks.  This is analogous to the $\langle \psi \psi \rangle$ propagator of superconductivity \cite{nam}.  To lowest order in this order parameter the diagonal propagators of the $\tilde{t}_{L,R}$ fields are just those already shown in the BS equation:
\begin{equation}
\label{tprop}
\Delta_{LL}(p)=\Delta_{RR}(p)=\frac{1}{p^2+M^2}
\end{equation}
while the $LR$ mixing propagator is
\begin{equation}
\label{mixprop}
\Delta_{LR}(p)=\frac{1}{p^2+M^2}\delta M^2(p^2)\frac{1}{p^2+M^2}.
\end{equation}

Away from the symmetry-breaking threshold, one would expect that for larger values of $A_t$ the mass of the Higgs particle moves away from zero, while the Nambu-Goldstone fields remain massless.
  This can proceed (for example, see \cite{kugo}) through a tachyonic solution to the BS equation, much as in the stabilization of a Mexican-hat potential where the stable non-perturbative vacuum yields a condensate and a Higgs boson of normal mass.  As for the Nambu-Goldstone bosons,
since the pioneering work of Nambu \cite{nam} we know that these massless Nambu-Goldstone excitations occur as a consequence of a non-trivial solution to a gap equation, an integral equation whose solution is a symmetry-breaking order parameter. This gap equation is essentially the BS equation at $P=0$.  (See \cite{delscad} for a proof of this Nambu-Goldstone mechanism in gauge theories such as QCD.)  We express the dynamics of symmetry breaking through the usual two-particle irreducible (2PI) effective potential $\Gamma$ \cite{cjt}, in which $\Gamma$ is a functional of $\delta M^2(p^2)$.
In this first investigation of the bound-state Higgs we ignore a number of interesting phenomena, including the VEV of the elementary Higgs fields and their possible mixing with the bound-state Higgs, so the effective potential (in the notation of \cite{cjt}) is 
\begin{equation}
\label{effpot}
\Gamma = \frac{1}{2}\mathrm{Tr} \{\ln G +[1-GG^{-1}_0]\} + \textrm{2PI graphs}, 
\end{equation} 
where the trace is over space-time as well as other relevant indices, such as particle type, $G$ is the exact propagator, and $G_0$ is the free propagator (when relevant; the term in square brackets is omitted for the $LR$ propagator).   The extrema of $\Gamma$ as the $G$ are varied yield the Schwinger-Dyson equations of the theory.

\begin{figure}
\includegraphics[width=0.5\textwidth]{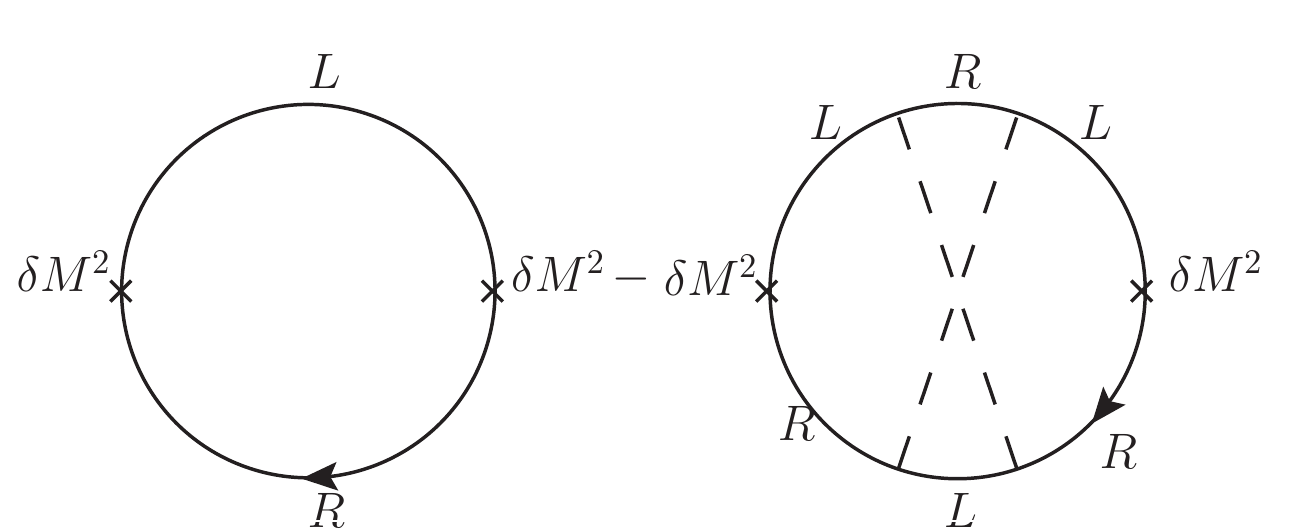}
\caption{\label{effpot2}  Diagrams for the effective potential quadratic in $\delta M^2$ (weights not shown).}
\end{figure}

To lowest order in $\delta M^2$ the effective action is given by the diagrams shown in Fig.~\ref{effpot2}, which give for $\Gamma$ the expression
\begin{equation}
\label{gammapart}
\Gamma = \frac{1}{2}\int \mathrm{d}^4p \rho (p^2) [\delta M^2(p^2)]^2-\frac{A_t^4}{2(2\pi )^4}\int \mathrm{d}^4p\,\int \mathrm{d}^4k \rho (p^2)\delta M^2(p^2) K(p,k)\rho (k^2)\delta M^2(k^2)
\end{equation} 
where
\begin{equation}
\label{rho}
\rho (k^2)=\frac{1}{(k^2+M^2)^2}.
\end{equation}

 Variation of the quadratic terms in $\Gamma$  with respect to $\delta M^2$ yields
\begin{equation}
\label{gapeq}
\delta M(p)^2 = \frac{A_t^4}{(2\pi )^4}\int d^4k \frac{1}{k^2+M^2}\delta M(k)^2\frac{1}{k^2+M^2}K(p,k).
\end{equation} 
This equation is analogous to standard gap equations for chiral symmetry breaking.  Just as for chiral gap equations it is in fact the original BS equation, in vertex form, at $P=0$, illustrating as before \cite{nam,delscad} the necessary existence of composite Nambu-Goldstone bosons when symmetries are broken without elementary Higgs fields.  It differs from chiral symmetry breaking gap equations because the kernel is well-behaved in the UV and there are no UV divergences.  The kernel falls like $1/p^4$ (modulo logarithms) at large momentum, implying the same falloff for $\delta M^2$, and Eq.~(\ref{gapeq}) is finite.

To analyze Eq.~(\ref{gapeq}) we need to analyze first the kernel $K(p,k)$. This kernel has the form
\begin{eqnarray}
\label{cross}
K(p,k) & = & \frac{1}{(2\pi )^4}\int d^4l\;\frac{1}{[l^2+M^2][(p+l)^2+m^2][(k+l-p)^2+M^2][(k+l)^2+m^2]}\\ \nonumber
& = & \frac{1}{16\pi^2}\int \prod dx_i \delta \left (1-\sum x_i \right )\frac{1}{D^2}
\end{eqnarray}
with
\begin{equation}
\label{denom}
D=k^2(x_1x_2+x_3x_4)+p^2(x_1x_4+x_2x_3)+(p+k)^2x_2x_4+(p-k)^2x_1x_3+M^2(x_2+x_4)+m^2(x_1+x_3).
\end{equation}
Now suppose that $M^2\gg m^2$, in which case $x_2,x_4$ have to be small compared to the other Feynman parameters.  So write $x_2=\lambda x, x_4=\lambda (1-x)$, with new integration variables running from 0 to 1. The integral over $\lambda$ will be dominated by small $\lambda$, so we can drop this variable judiciously.  Then approximately 
\begin{equation}
\label{newvar}
\prod dx_i \delta \left (1-\sum x_i \right )=\lambda d\lambda dx dx_1dx_3 \delta (1-x_1-x_3)
\end{equation}
and
\begin{equation}
\label{approx}
D= x_1x_3(p-k)^2+m^2(x_1+x_3)+\lambda [ak^2+(1-a)p^2+M^2]
\end{equation}
where
\begin{equation}
\label{abc}
a=x_1x+x_3(1-x),\;0\leq a \leq 1,
\end{equation}
and we dropped a term $\sim \lambda^2$ in $D$.

Now do the integral over $\lambda$ explicitly, with the result     
\begin{equation}
\label{doint}
K(p,k)\approx \frac{1}{16\pi^2}\int dx_1dx_3dx\delta (1-x_1-x_3)  \left \{\frac{1}{A^2}\ln \left [\frac{A+B}{B} \right ]-\frac{1}{A(A+B)} \right \}
\end{equation}
where
\begin{equation}
\label{candd}
A= [ak^2+(1-a)p^2+M^2],\;B=x_1x_3(p-k)^2+m^2.
\end{equation}

We now show  that at large $k^2\gg p^2$, $K\sim (1/k^4)\ln k^2$ and so vanishes rapidly.  This means that the integral in the gap equation also vanishes rapidly, as we will soon see. 
Since $B$ does not depend on $x$  the integral over $x$ can be done, yielding
\begin{equation}
\label{largek}
K\rightarrow \frac{\ln k^2}{8\pi^2  k^4},
\end{equation}
a result that we can get exactly at $p=0$. 

The next step is to reduce the gap equation to a one-dimensional equation by integrating over the angles of $k$.
In the gap equation, because the angle between the four-momenta  appears only in logarithms or in a term parametrically 
small with respect to $M^2$, we make the approximation
\begin{equation}
\label{angint}
\int d\Omega_k F[(p-k)^2]\approx 2\pi^2[\theta (p^2-k^2)F(p^2)+\theta (k^2-p^2)F(k^2)].
\end{equation}  
This is exactly true for $F=1/(p-k)^2$ or for constant $F$, and is acceptable for the logarithmic functions we encounter.    
After projecting out the s-wave, to lowest (quadratic) order in $\delta M^2$ the relevant part of $\Gamma$ is
\begin{equation}
\label{gammapart2}
\Gamma = \frac{1}{2}\int \mathrm{d}p^2\,p^2\, \rho (p^2) [\delta M^2(p^2)]^2-\frac{A_t^4}{2(2\pi )^4}\int \mathrm{d}p^2\,p^2\,\int \mathrm{d}k^2\,k^2\, \rho (p^2)\delta M^2(p^2) \hat{K}(p^2,k^2)\rho (k^2)\delta M^2(k^2)
\end{equation} 
where $\hat{K}$ is the s-wave projection of the kernel.
 
 Variation of this equation yields the s-wave projection of Eq.~(\ref{gapeq}), which becomes a standard one-dimensional homogeneous Fredholm integral equation with a discrete spectrum of eigenvalues $A_t^4$. We seek the lowest eigenvalue by inserting a trial function into Eq.~(\ref{gammapart2}) and doing the integrals numerically, including an approximation to the integral over
$x_1$  in the kernel (see Eq.~(\ref{newvar})), for various values of the Higgs-to-stop mass ratio $m/M$.

\begin{figure}[ht!]
\includegraphics[width=0.4\textwidth]{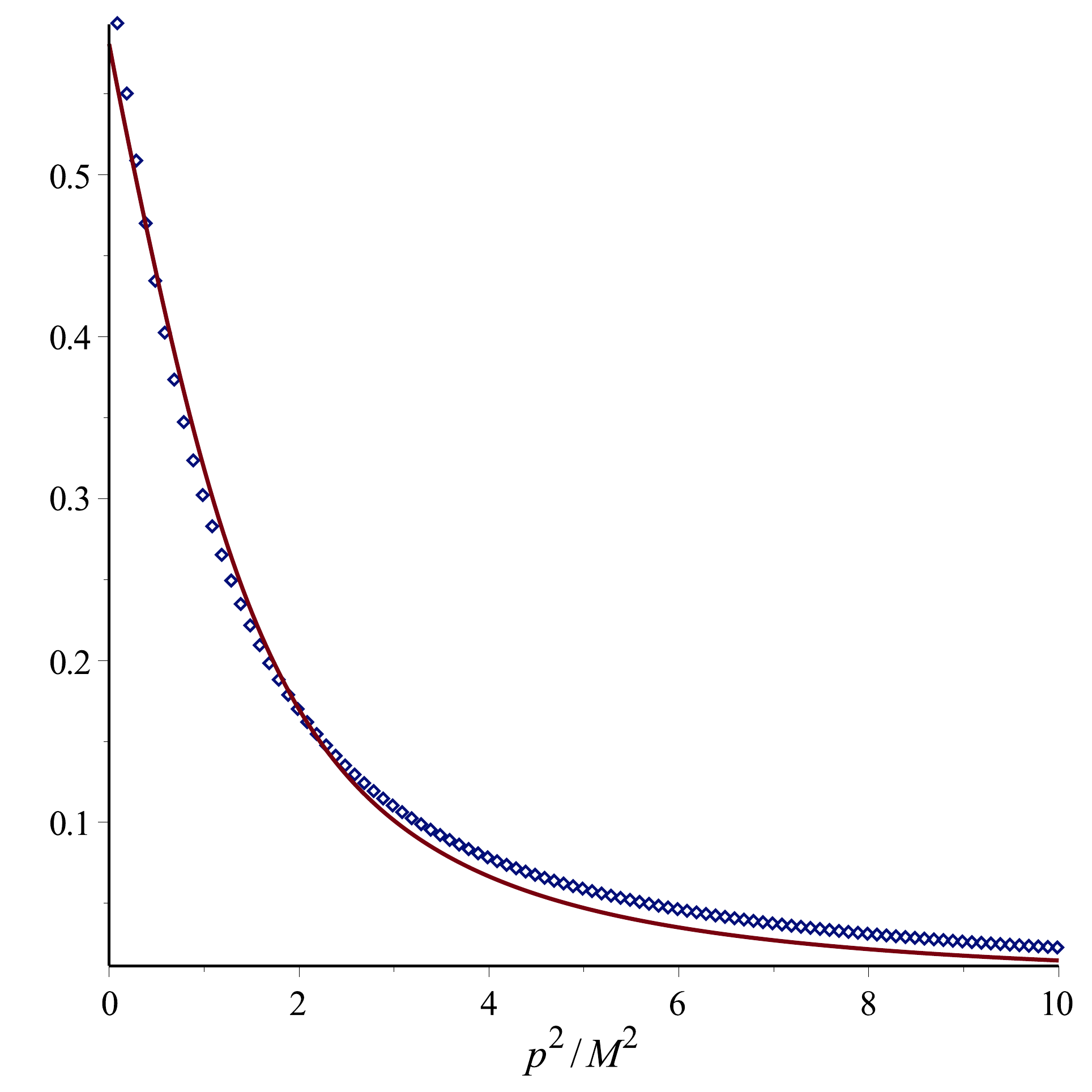}
\caption{\label{fig4-eigen} A comparison of the input and output using the second-order eigenfunction $\delta M^2_2$, as a function of $p^2/M^2$, calculated numerically as   described in the text.  In the case of the exact solution, the two curves would be identical.}
\end{figure}

We motivate our trial functions for the crossed-graph kernel of interest from known exact results \cite{wick} for  the BS equation with the massless kernel
\begin{equation}
\label{wickker}
K\sim \frac{1}{(p-k)^2}.
\end{equation} 
For the vertex form of the BS equation, given in Eq.~(\ref{bseq}), the lowest eigenfunction at $P=0$ is
\begin{equation}
\label{wickeigen}
Q(0,p) \equiv \delta M^2(p^2)\sim \frac{1}{p^2+M^2}.
\end{equation}
Naturally, the class of trial functions of this form with   $M^2$ replaced by a variational parameter $\mu^2$ yields the exact result.  In the present problem the asymptotic behavior is different, so we choose as a zeroth-order trial function
\begin{equation}
\label{oureigen}
\delta M^2_0(p^2)\sim \frac{1}{p^4+\mu^4}.
\end{equation}
We have studied other trial functions, such as $1/(p^2+\mu^2)^2$, with similar results.  We improve this first variational estimate by using $\delta M^2_0$ as input to the right-hand side of Eq.~(\ref{gammapart2}), numerically calculating a new output $\delta M_1^2$. 
We made a simple but accurate fit to $\delta M^2_1$, amounting to adding a term $\sim p^2$ to the denominator of Eq.~(\ref{oureigen}).  
Then we used the average $\delta M^2_2\equiv (1/2)(\delta M^2_0+\delta M^2_1)$ as a trial function, and calculated the output again.  
This yields excellent agreement between the new input and output, as shown in Fig.~\ref{fig4-eigen}, for the specific value $m/M$ = 0.05.

At this elementary Higgs mass the   critical coupling resulting from our numerical calculations is
\begin{equation}
\label{minat}
\frac{A_t}{M}\approx 15.14.
\end{equation}
This  estimate  is to be compared with the value of $A_t/M=\sqrt{6}$ that maximizes the lightest Higgs mass in an approximate one-loop calculation~\cite{dmrs,Haber:1996fp}.  But this is not the final verdict, since mixing of the bound Higgs with the MSSM Higgs and other possible bound states need investigation.  

\begin{figure}
\includegraphics[width=0.4\textwidth]{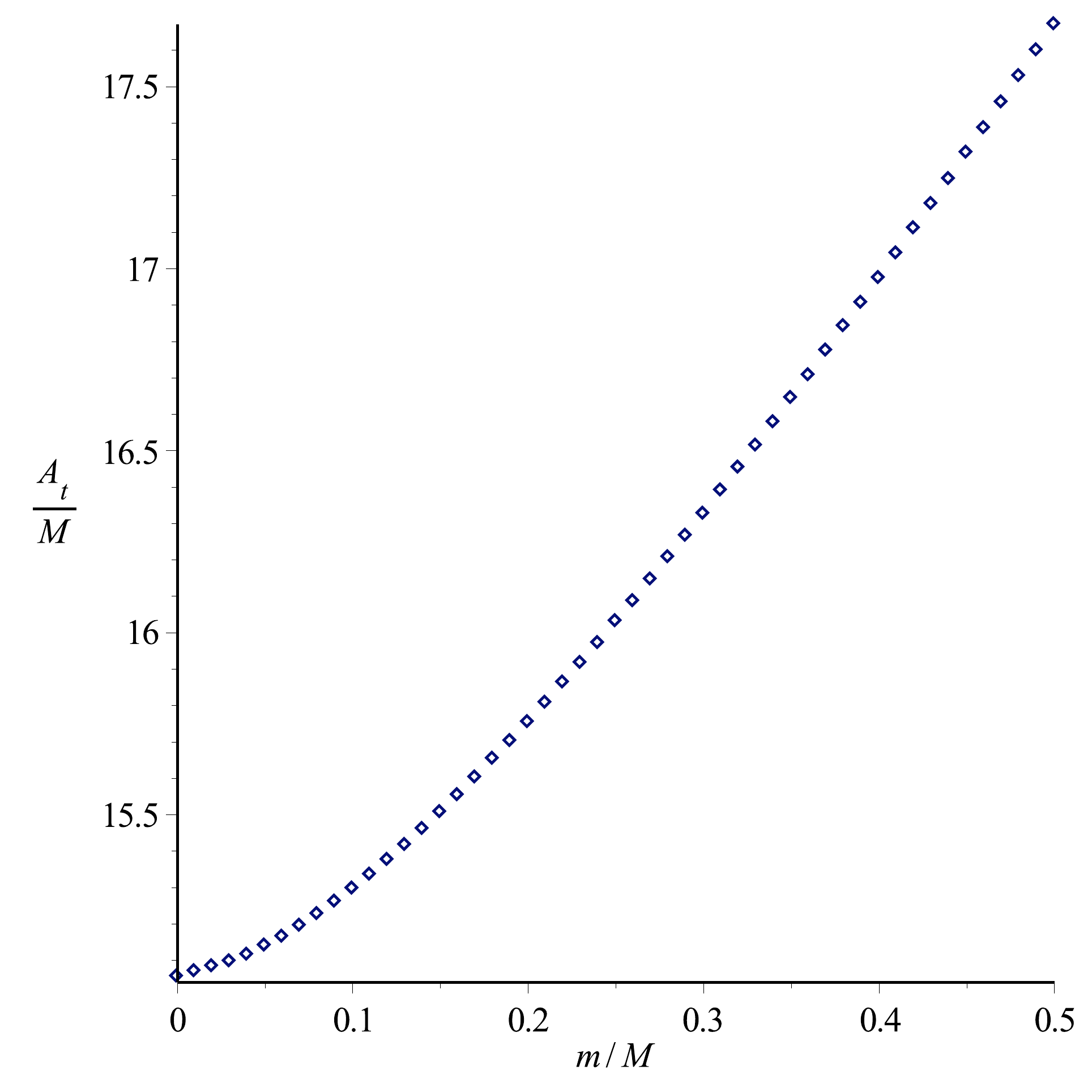}
\caption{\label{fig5-coup-mass} Behavior of the critical coupling as a function of the elementary Higgs mass.}
\end{figure}

Also of interest is the needed critical coupling for various masses of the elementary  Higgs field.  This is shown in
Fig.~\ref{fig5-coup-mass}.  As expected, the critical coupling increases with increasing Higgs mass.

While it appears plausible that supersymmetry breaking in the MSSM can trigger electroweak symmetry breaking via the formation of bound states with non-zero VEVs, further work is needed before one can build a realistic model and compare its predictions with the data.   In addition to the color-singlet states, the same trilinear scalar interactions can cause  colored bound states to form.  The viability of the model depends on its ability to produce a standard-model-like vacuum with broken SU(2)$\times$U(1) but unbroken SU(3), in which case the colored bound states are turned into color singlets by strings attached to gluons or quarks.  It is well known that the MSSM, in its traditional realization, has a number of dangerous color and charge breaking minima, although  cosmological evolution favors the vacuum with unbroken SU(3) even in some cases where it is not the global minimum of the potential~\cite{Kusenko:1996jn}.  In our case, one must now re-examine the same issue taking into account a number of new effective degrees of freedom.  While the full analysis is obviously complicated and the results will inevitably be model-dependent, there is one feature of the color-singlet states that sets them apart from the rest.   The color-singlet states can have a mixing with the fundamental Higgs bosons via the same coupling as that which enters the BS equation.  The mass matrix in the bound-state--Higgs basis has both  diagonal terms and  off-diagonal terms.  In contrast, the colored bound states can only have diagonal terms.  As the BS coupling increases, the mass squared of each bound state decreases, as discussed above. Since the scalar exchange forces are essentially color-blind, the bound states with different SU(3) properties can have similar binding energies.  However, thanks to the off-diagonal terms, the colorless bound states can develop a VEV simultaneously with the Higgs boson for some value of the trilinear coupling for which the diagonal terms are still positive.  This possibility leads to an appropriate standard vacuum.

\begin{figure}[ht!]
\raisebox{0.5\height}{\includegraphics[width=0.45\textwidth]{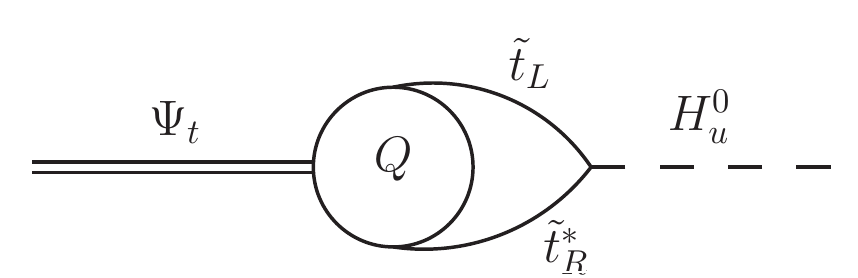}}  
\includegraphics[width=0.35 \textwidth]{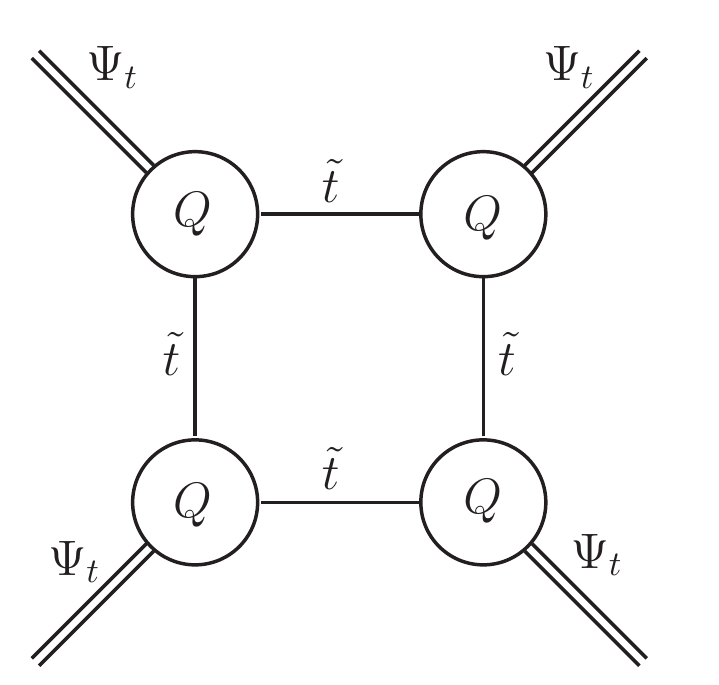}
\caption{\label{mixing} Mixing of the composite and elementary Higgs fields, and four-point couplings of the composite Higgs field.}
\end{figure}

The mixing of the fundamental Higgs and the bound state occurs through a diagram involving the solution to the BS equation, as shown in Fig.~\ref{mixing}.  Also shown is an approximation to the four-point coupling of the composite Higgs, again expressed in terms of the BS solution.  This quartic coupling is not related to the gauge couplings by supersymmetry because it depends explicitly on the supersymmetry breaking parameter $A_t$.  Therefore, in the vacuum where the fundamental Higgs boson and the bound state mix, the usual upper bound on the lightest Higgs boson mass does not apply.  

Electroweak symmetry breaking occurs via a {\em symmetry-breaking seesaw}, when the following mass matrix has a negative eigenvalue: 
\begin{equation}
{\cal M}_{\rm H}^2 = \left( \begin{array}{ll}
         \mu_{\rm H}^2 & M_{\rm HB}^2 \\
          M_{\rm HB}^2 & M_{\rm B}^2
        \end{array}
\right ).
\end{equation}
Here $\mu_{\rm H}^2$ is the Higgs mass-squared parameter, which is negative in the Standard Model, but which we take to be positive; $ M_{\rm HB}$ is the mixing parameter calculated from the diagram shown in Fig.~\ref{mixing}, and $M_{\rm B}$ is the mass of the bound state.  Our calculations above dealt with $M_{\rm B}^2=P^2=0$,  which required the large value of $A_t/M$ quoted in equation (\ref{minat}).  However, in the presence of mixing, the symmetry breaking can occur for a non-zero $M_{\rm B}$ and a positive $\mu_{\rm H}^2$, as long as 
\begin{equation}
\det {\cal M}_{\rm H} = \mu_{\rm H}^2 M_{\rm B}^2 - M_{\rm HB}^4 \, < \, 0.
\end{equation}

The value of the mixing parameter $M_{\rm HB}$ can be extracted from a solution of the 
BS equation with a proper wave function normalization. We leave this calculation for future work.  Here we parameterize  
\begin{equation}
M_{\rm HB} = A_t \xi,  
\end{equation}
where $\xi$ is dimensionless number (which may depend on $A_t$ and on other parameters).   Now the critical value of $A_t$ is different from that in Eq.~\ref{minat}: 
\begin{equation}
 \frac{A_t}{M} = \min \left \{ \sqrt{\frac{1}{ \xi} \frac{\mu_{_{\rm H}}}{M}}, \ 15.4 \right \}.
\end{equation}

Given that $\mu_{\rm H}$ can be arbitrarily close to zero (and, in fact, $\mu_{\rm H}^2<0$ in the Standard Model), the symmetry breaking can occur even for smaller values of $A_t/M$.  Of course, if $\mu_{\rm H}^2$ has to be fine-tuned to be small, the scenario in question becomes rather contrived.  For reasonable values of parameters and for the binding energy that is not as large as the mass, which is the case for a smaller $A_t$, there is a self-consistent set of values, for example, 
\begin{equation}
\mu_{_{\rm H}}\sim 10^2 {\rm \, GeV}, \ \ M_{\rm B} \sim 2M \sim 2 \, {\rm TeV}, \ 
A_t\sim \sqrt{\frac{1}{ \xi} \mu_{_{\rm H}} M_{\rm B}} \sim  300 {\rm \, GeV} \sqrt{\frac{1}{ \xi} }, 
\end{equation}
and, thus,
\begin{equation}
 \frac{A_t}{M} \sim 0.3  \sqrt{\frac{1}{ \xi} }.  
\label{AtoverM}
\end{equation}

One can now ask whether these values are consistent with vacuum stability with respect to dangerous color and charge breaking minima in the potential~\cite{ccb,Kusenko:1996jn}.  In general, one has to analyze the full potential of the MSSM to identify the global and the local minima.  
If the global minimum is color and charge preserving, the model is viable.  If the standard-model-like vacuum is only a local minimum, one has to examine cosmological evolution because in many cases the vacuum that is populated first is the one in which the squarks have zero VEV~\cite{Kusenko:1996jn,Kusenko:1996xt}.  One then has to calculate the lifetime of the false vacuum and determine whether the universe is likely to undergo cooling and remain in the false vacuum until the present time.  An approximate criterion for the absence of dangerous color and charge breaking minima, based on the analysis of Ref.~\cite{Kusenko:1996jn}, is 
\begin{equation}
 \frac{A_t}{M} \lesssim 2.7, 
\end{equation}
which can be consistent with Eq.~(\ref{AtoverM}) in some reasonable range of parameters.  

A large trilinear coupling at low energy can become even larger at higher scales, but the effects of renormalization group running depend on the size of the coupling.  For negative sign, the initially large $A_t$ can cross zero at a high energy scale~\cite{dmrs}.

In addition to the SU(2) doublet bound state, the large $A_t$ can lead to formation of a gauge singlet state in which $\{ {\tilde t_L},\ {\tilde t_L}^* \}$ in the initial state exchange the Higgs boson and convert into $\{ {\tilde t_R},\  {\tilde t_R}^* \}$, which, in turn exchange the Higgs boson and 
convert back to $\{ {\tilde t_L},\ {\tilde t_L}^* \}$.  This is similar to the doublet bound state we have considered, but with a BS wavefunction  appropriate for the one-particle kernel.  The existence of such a gauge singlet scalar is a generic prediction of our model.\footnote{We thank T.T.~Yanagida for pointing this out, and for other valuable comments.} 

So far, we have focused on the version of MSSM in which only the top trilinear term is large.  However, if the other trilinear couplings and the 
$\mu B$ term are  large as well, new additional states can form through the Higgs boson exchange, and new diagrams can contribute to the bound states we  discussed. 

The electroweak precision measurements should impose constraints on any strongly coupled model built along the lines we discussed.  These constraints imply a lower bound on the mass of the squarks.  It is possible, and, in fact, likely, that the squark masses would have to be of the order of $5-10$~TeV for the model to be consistent with the precision measurements.  This forces the $A$ term to be correspondingly larger, and there may appear to be a small hierarchy between the supersymmetry breaking scale and the electroweak scale.  This hierarchy may impose some degree of fine-tuning on a realistic model based on  strongly coupled broken supersymmetry.  This is a likely potential drawback of the otherwise very appealing scenario, in which the scale of the electroweak symmetry breaking is determined by the breaking of supersymmetry.  However, the class of models we discuss still possesses a robust solution to the big hierarchy problem: above the scale of the bound states, the MSSM exists in its usual incarnation, and supersymmetry stabilizes the scales in the usual way.  

We have examined the possibility of electroweak symmetry breaking by the formation of bound states of squarks via the Higgs boson exchange, which bound states can mix with the fundamental Higgs bosons and can acquire VEVs simultaneously with these fundamental  bosons.  This scenario is clearly different from the widely discussed technicolor models~\cite{technicolor}, including walking technicolor~\cite{Akiba:1985rr}, models with color singlets~\cite{Ryttov:2011my}, and the models in which supersymmetry and technicolor are combined~\cite{Dine:1981za}.  
Our scenario has the potential to relate the scales of supersymmetry breaking and electroweak breaking in a new way, but the applications to realistic models requires a more detailed analysis.  It is evident that 
bound states can play an important role, and they should not be neglected in models with large trilinear supersymmetry breaking terms.  The next step that we plan to undertake is to investigate the mixing of the elementary and composite Higgs bosons and the resulting symmetry-breaking patterns.  These results will be presented elsewhere.  

\acknowledgments

We thank T.T.~Yanagida for stimulating discussions and for many valuable comments.
This work was supported in part by DOE Grant DE-FG03-91ER40662.  A.K. appreciates support of World Premier International Research Center Initiative, MEXT, Japan, as well as hospitality of the Aspen Center for Physics, which is supported by the NSF Grant No. PHY-1066293.

\end{document}